\documentclass[aps,prb,twocolumn,groupedaddress,amsmath]{revtex4}
	\usepackage{bm}
	\usepackage{graphicx}
\newcommand\beq{\begin{equation}}
\newcommand\eeq{\end{equation}}
\newcommand\beqa{\begin{eqnarray}}
\newcommand\eeqa{\end{eqnarray}}

\begin{document} 
\title{Contact values of the {radial} distribution functions of additive 
hard-sphere mixtures in $d$ dimensions: A new proposal} 
\author{Andr\'es Santos}
\email[]{andres@unex.es}
\author{Santos B. Yuste} 
\email[]{santos@unex.es}
\affiliation{Departamento de F\'{\i}sica, Universidad de Extremadura, Badajoz, 
E-06071, Spain}
\author{Mariano L\'{o}pez de Haro} 
\email[]{malopez@servidor.unam.mx}
\affiliation{Centro de Investigaci\'on en Energ\'{\i}a, UNAM, Temixco, 
Morelos 62580, M{e}xico} 

\date{\today}

\begin{abstract} 
The contact values $g_{ij}(\sigma_{ij})$ of the radial distribution functions of a $d$-dimensional mixture of {(additive)} hard spheres are considered. 
{A `universality' assumption is put forward, according to which 
$g_{ij}(\sigma_{ij})=G(\eta, z_{ij})$, where $G$ is a common 
function for all the mixtures of the same dimensionality, regardless of the 
number of components, $\eta$ is the packing fraction of the mixture, and 
$z_{ij}=(\sigma _{i}\sigma _{j}/\sigma _{ij})\langle \sigma^{d-1}\rangle 
/\langle \sigma ^{d}\rangle$
is a dimensionless parameter, $\langle \sigma^n\rangle$ being the $n$-th moment of the diameter distribution. For $d=3$,} this universality assumption holds 
for the contact values of the Percus--Yevick approximation, the Scaled 
Particle Theory, and, {consequently,} the 
{Boubl\'{\i}k--}Grundke--Henderson--Lee--Levesque approximation.
{Known exact consistency conditions are used to express $G(\eta,0)$, $G(\eta,1)$,
and $G(\eta,2)$ in terms of the radial distribution at contact of the one-component
system. Two specific proposals consistent with the above conditions (a 
quadratic form and a rational form) are made for the $z$-dependence of $G(\eta,z)$.}
For one-dimensional systems, the proposals for the 
contact values reduce to the exact result. Good agreement between the
predictions of the proposals and available numerical results is found 
for $d=2$, $3$, $4$, and $5$. 
\end{abstract}

\maketitle

\section{Introduction}
\label{sec1}

It is well known that there exists a close connection between 
the
thermodynamic and structural properties of classical
fluids. In fact, for 
pair-wise additive intermolecular potentials, all the
thermodynamic 
functions may be expressed in
terms of the radial distribution functions 
(rdf). The expressions are
particularly simple for hard-core fluids, since 
in
that case the internal energy reduces to that of the ideal gas and in 
the
pressure equation it is only the contact values
rather than the full rdf 
which appear explicitly. Therefore, knowledge of
the contact values of the 
rdf in hard-core
fluids, which we will denote by $g_{ij}(\sigma _{ij})$ 
(where in general
$i$ and $j$ label species and $\sigma _{ij}$ is
the 
distance of separation at contact between the centers of two
interacting 
fluid particles, one of species $i$ and the
other of species $j$), suffices 
to obtain the equation of state (EOS) of
these systems. In the case of a 
single component
hard-core fluid, if the EOS were known then it would be 
straightforward to
infer the contact value of the radial
distribution 
function. In contrast, if one were given the EOS of a
multicomponent 
hard-core mixture such contact values
could not be determined in a unique 
way. Up to this day, no exact
expressions neither for the contact values of 
the rdf
nor for the EOS (except for the case of one-dimensional systems, 
{ i.e.}, hard rods) are known, although various
approximate theories, 
empirical efforts, and computer simulations have been
carried out in 
connection with this problem.
One-component systems are of course easier to 
handle and this has meant
that studies for mixtures are much 
scarcer.
Perhaps the most successful theoretical (analytical) approach to 
this issue
in the case of  additive hard-sphere mixtures 
($d=3$) is the exact solution of the Percus--Yevick (PY)
equation carried 
out by Lebowitz in 1964. \cite{L64} This analytical solution, which among 
other things yields explicit
expressions for the contact values of the rdf, 
{as well as for} 
the virial and the compressibility routes to the 
EOS, is at the
basis of the celebrated (and empirically
derived) 
Boubl\'{\i}k--Mansoori--Carnahan--Starling--Leland (BMCSL) EOS, 
\cite{B70,MCSL71}
considered to be a rather accurate EOS
for hard-sphere mixtures. 
{In his paper, \cite{B70} Boubl\'{\i}k also introduced an
approximation for the contact values of the
rdf (in fact an 
interpolation between the PY results and the ones of the
Scaled Particle 
Theory (SPT) \cite{LHP65,R88}) that later was independently proposed by 
Grundke and Henderson \cite{GH72} and Lee and Levesque. \cite{LL73} This 
approximation,
which we will refer to as BGHLL, leads precisely to the BMCSL EOS  when 
substituted into the statistical mechanical formula for the pressure 
equation.} Refinements of the BGHLL approximation have been recently proposed by  Henderson and Chan,  \cite{HMLC96,YCH96,YCH97,HC98,MHC99} Matyushov and Ladanyi, \cite{ML97}
and Barrio and Solana \cite{BS00} to cope with some deficiencies of the 
BMCSL EOS in the so-called colloidal limit of {binary} hard-sphere 
mixtures.

As far as we are aware, there are no reported approximations 
for $g_{ij}(\sigma _{ij})$ {with $d\neq 3$, except that of 
Jenkins and Mancini 
\cite{JM87}
in the case of hard-disk mixtures and our 
own \cite{SYH99} for $d$-dimensional mixtures.
The
latter  approximation, 
however, was introduced only as a means to
derive a proposal for the EOS of mixtures. In fact, while this EOS presents 
an excellent agreement with simulations
for $d=2$, \cite{SYH99,HYS02} $d=3$, \cite{SYH99,MV99,CCHW00} $d=4$, \cite{GAH01}
and $d=5$, \cite{GAH01} the expressions for $g_{ij}(\sigma _{ij})$ are less accurate. 
\cite{CCHW00,GAH01}}
 It is the major aim of this paper to propose new (improved)
approximate expressions for $g_{ij}(\sigma _{ij})$, for 
arbitrary mixtures and arbitrary dimensionality, that, apart from
satisfying  known consistency conditions, are
sufficiently 
general and flexible to accommodate any given EOS for the single fluid. A key aspect 
of the present approach, also included in our previous work, 
\cite{SYH99} is that we will assume from the very beginning a kind 
of universal behavior of the contact values which at 
least holds also for the  solution of the PY equation,\cite{L64} for the 
SPT approximation, \cite{LHP65,R88} and for the BGHLL 
interpolation \cite{B70,GH72,LL73} in the case of mixtures of hard spheres 
$(d=3)$.
This means that, once the dimensionality  and the packing fraction are
fixed, the expression for the contact values $g_{ij}(\sigma_{ij})$ for
all pairs $ij$ is the same, irrespective of the composition and the number of components in the
mixture. 
This expression must comply with  three
consistency conditions related to the point particle, equal size, and the
colloidal limits, respectively.
Two functional forms (a quadratic one and a
rational one) which are sufficiently representative  will be examined. Their merits will be
assessed from a comparison with available simulation data as well as with
respect to the performance of the EOS obtained from them. In this latter
issue, we will show that a paradoxical result is obtained. What we find is
that, contrary to what one could possibly expect, better contact values
{\em do not\/} necessarily mean {\em more accurate\/} EOS and that even two
{\em different\/} expressions for $g_{ij}(\sigma_{ij})$  may lead to
exactly the {\em same\/} EOS.

The paper is organized as follows. In Sec.\ \ref{sec2} we recall the
known consistency conditions and introduce the new proposals for the
contact values of the rdf. Section \ref{sec3} deals with the comparison
between our contact values and ensuing EOS and simulation results. We
close the paper in Sec.\ \ref{sec4} with further discussion and some
concluding remarks.

\section{The proposal\label{sec2}}
Let us consider a mixture of hard spheres in $d$ dimensions with 
an
arbitrary number of components. The hard core of the
interaction between 
a sphere of species $i$ and a sphere of species $j$ 
is
$\sigma_{ij}=\frac{1}{2 }(\sigma _{i}+\sigma _{j})$, where the diameter 
of
a sphere of species $i$ is
$\sigma _{ii}=\sigma _{i}$. The number density 
of the mixture is $\rho $
and the mole fraction of species $i$ 
is
$x_{i}=\rho _{i}/\rho $. {}From these quantities one can define the 
packing
fraction
$\eta =v_{d}\rho \langle \sigma ^{d}\rangle $, 
where
$v_{d}=(\pi /4)^{d/2}/\Gamma (1+d/2)$ is the volume of a 
$d$-dimensional sphere of unit diameter and
$\langle \sigma ^{n}\rangle 
\equiv \sum_{i}x_{i}\sigma _{i}^{n}$ denotes
the
moments of the diameter 
distribution.

In a hard-sphere mixture, the knowledge of the contact 
values
$g_{ij}(\sigma_{ij})$ is important for a number of
reasons. For 
example and as stated above, the availability of
$g_{ij}(\sigma _{ij})$ is 
sufficient to get the equation of
state (EOS) of the mixture via the virial 
expression 
\begin{equation} 
Z_{\text{m}}(\eta )=1+2^{d-1}\eta 
\sum_{i,j}x_{i}x_{j}\frac{\sigma _{ij}^{d} }{\langle \sigma 
^{d}\rangle}g_{ij}(\sigma_{ij}),  
\label{1} 
\end{equation} 
where
$Z_{\text{m}}=p/\rho k_{B}T$ is the compressibility factor of the 
mixture,
$p$ being the pressure, $k_{B}$ the Boltzmann
constant, and $T$ the 
absolute temperature. The contact values
$g_{ij}(\sigma _{ij})$ are also 
needed to generate the
entire rdf $g_{ij}(r)$ in the Generalized Mean 
Spherical Approximation
\cite{GAC85} and in the Rational 
Function
Approximation. \cite{YSH98} In a different context, they are 
important as
well to implement the Enskog kinetic theory
both for elastic 
and inelastic hard spheres. \cite{JM87,GD99}

The exact form of $g_{ij}(\sigma _{ij})$ as functions of the 
packing
fraction, the set of diameters $\{\sigma _{k}\}$,
and the set of 
mole fractions $\{x_{k}\}$ is only known in the one-dimensional case, where 
one simply has $g_{ij}(\sigma _{ij})=(1-\eta )^{-1}$. 
Consequently, {for $d\geq 2$} one has to resort to
approximate 
theories or empirical expressions.
{}From that point of view, it is useful 
to make use of exact limit results
that can help one in the construction 
of
approximate expressions for $g_{ij}(\sigma _{ij})$. Let us consider 
first
the limit in which one of the species, say $i$,
is made of point 
particles that do not occupy volume, i.e.,
$\sigma _{i}\rightarrow 0$. In 
that case,
$g_{ii}(\sigma _{i})$ takes the ideal gas value, except that the 
available
volume fraction is $1-\eta$: \begin{equation}
\lim_{\sigma_{i}\rightarrow 
0}g_{ii}(\sigma_{i})=\frac{1}{1-\eta}.
\label{2} 
\end{equation} 
An even simpler situation
occurs when all the species have the same 
size,
$\{\sigma _{k}\}\rightarrow \sigma $, so that the system 
becomes
equivalent to a one-component system. Therefore, 
\begin{equation} 
\lim_{\{
\sigma _{k}\}\rightarrow \sigma }g_{ij}(\sigma_{ij})=g(\sigma ), 
\label{3} 
\end{equation} 
where $g(\sigma )$ is the contact value of the radial distribution 
function in the one-component case. Equations (\ref{2}) and (\ref{3}) 
represent the simplest and most basic conditions that $g_{ij}(\sigma _{ij})$ 
must satisfy. There is a number of other {less trivial} consistency 
conditions. \cite{R88,HMLC96,HC98,ML97,BS00,H94,HBCW98,V98,THM99} Here we consider 
the condition stemming from a binary mixture in which one of the species, 
say $i=1$, is much larger than the other one, i.e., $\sigma _{1}/\sigma 
_{2}\rightarrow \infty $, but occupies a
negligible volume, i.e., 
$x_{1}(\sigma _{1}/\sigma _{2})^{d}\rightarrow 0$. In that case, a sphere of 
species 1 is seen
as a wall by particles of species 2, so that 
\cite{HMLC96,HBCW98,RDA01}

\begin{equation} 
\lim_{\stackrel{\sigma _{1}/\sigma _{2}\rightarrow \infty}{ 
x_{1}(\sigma _{1}/\sigma _{2})^{d}\rightarrow 0}}\left[g_{12}(\sigma 
_{12})-2^{d-1}\eta g_{22}(\sigma_{2})\right] =1.  
\label{4} 
\end{equation} 
Also in that limit, \cite{HMLC96,HBCW98,RDA01} $\ln g_{11}(\sigma _{1})\sim 
\sigma _{1}/\sigma _{2}$, but we will not make use of this condition here.

Our purpose now is to propose approximate expressions for $g_{ij}(\sigma _{ij})$ of hard-core mixtures with an arbitrary number of 
components and arbitrary dimensionality $d$, that
satisfy the consistency 
conditions (\ref{2})--(\ref{4}). First, we assume that the dependence of 
$g_{ij}(\sigma_{ij})$ on the parameters $\{\sigma _{k}\}$ and
$\{x_{k}\}$ 
takes place \textit{only} through the scaled quantity
$z_{ij}\equiv (\sigma 
_{i}\sigma_{j}/\sigma _{ij})\langle \sigma^{d-1}\rangle /\langle \sigma 
^{d}\rangle$. More
specifically, 
\begin{equation} g_{ij}(\sigma 
_{ij})=G(\eta,z_{ij}),
\label{5} 
\end{equation} 
where the function $G(\eta,z)$ is \textit{universal} in the sense that it is a common function
for all 
the pairs $ij$, regardless of the {composition and} number of 
components of the mixture. Of course, the function $G(\eta ,z)$
is different 
for each dimensionality $d$. 

{The ratio $\xi\equiv 
\langle\sigma^{d-1}\rangle/\langle\sigma^{d}\rangle$ can be understood as a 
`typical' inverse diameter or curvature of the particles of the mixture. 
The
parameter $z_{ij}^{-1}=(\sigma_i^{-1}+\sigma_j^{-1})/2\xi$ represents then the average 
curvature, in units of $\xi$, of a particle of species $i$ and a particle of species $j$. 
According to Eq.\ (\ref{5}), if two different pairs $ij$ in two different mixtures 
(with the same packing fraction) have the same dimensionless average curvature $z_{ij}^{-1}$, 
then they have the same contact value of the rdf.}

The ansatz (\ref{5}) includes the one used by us, \cite{SYH99} where
$G(\eta 
,z)$ was approximated by a linear function of $z$. {A particular case 
of this linear form is the proposal made by Jenkins and Mancini for 
hard-disk mixtures: \cite{JM87}} 
\beq
G(\eta,z)=\frac{1}{1-\eta}+\frac{9}{16}\frac{\eta}{(1-\eta)^2}z, \quad (d=2).
\label{JM}
\eeq
In the three-dimensional case, 
Eq.~(\ref{5}) is also compatible with the solution of the PY equation, 
\cite{L64} the SPT approximation, \cite{LHP65,R88} and, consequently, the 
BGHLL
proposal. \cite{B70,GH72,LL73} More specifically, $G(\eta ,z)$ is a 
linear function of $z$ in the PY approximation and a
quadratic function in 
the SPT and BGHLL approximations:
\begin{equation} 
G(\eta ,z)=\frac{1}{1-\eta }+\frac{3}{2}\frac{\eta }{(1-\eta )^{2}}z+\lambda 
\frac{\eta^{2}}{(1-\eta)^{3}}z^{2},\quad (d=3),  
\label{15} 
\end{equation} 
where $\lambda_{\text{PY}}=0$, $\lambda _{\text{SPT}}=\frac{3}{4}$, and $ 
\lambda_{\text{BGHLL}}=\frac{1}{2}$. These three approximations 
are consistent with (\ref{2}) and (\ref{3}), but only the SPT is also consistent 
with condition (\ref{4}). The approximation referred to as the SPT-virial 
route by Rosenfeld \cite{R88} adopts also the scaling form (\ref{5}), 
namely
$G(\eta,z)=(1-\eta )^{-1}\exp \left[ 3z\eta /2(1-\eta )\right] $, 
{but it does not comply with condition (\ref{4})}.

Once we adopt the ansatz (\ref{5}), the limits in 
(\ref{2})--(\ref{4}) provide very useful constraints on the $z$-dependence 
of $G$. First, $ z_{ii}\rightarrow 0$ in the limit
$\sigma _{i}\rightarrow 
0$, so that insertion of Eq.~(\ref{2}) into (\ref{5}) yields 
\begin{equation} 
G(\eta ,0)=\frac{1}{1-\eta }.
\label{6}
\end{equation} 
Next, if all the diameters are equal, $z_{ij}\rightarrow 1$, so that 
Eq.~(\ref{3}) implies that 
\begin{equation} 
G(\eta ,1)=g(\sigma ).
\label{7} \end{equation} 
Finally, in the limit 
considered in Eq.~(\ref{4}),
we have $ z_{22}\rightarrow 
1$, $z_{12}\rightarrow 2$. Consequently, 
\begin{equation} G(\eta ,2)=1+2^{d-1}
\eta G(\eta ,1).  
\label{8} 
\end{equation} 
Thus {Eqs.\ (\ref{6})--(\ref{8}) provide  complete information on 
the 
function $G$ at $z=0$, $z=1$, and $z=2$, in terms of the contact value 
$g(\sigma)$ of the one-component rdf.}

The proposal made in Ref.~\onlinecite{SYH99} consists of assuming a 
linear dependence of $G$ on $z$ that satisfies the
requirements (\ref{6}) 
and (\ref{7}): 
\begin{equation} 
G(\eta,z)=\frac{1}{1-\eta }+\left[ g(\sigma 
)-\frac{1}{1-\eta }
\right] z. 
\label{9} 
\end{equation} 
{If in the two-dimensional case we take Henderson's value \cite{H75}
$g(\sigma)=(1-7\eta/16)/(1-\eta)^2$, Eq.~(\ref{9}) reduces to Jenkins and 
Mancini's approximation,\cite{JM87}} Eq.~(\ref{JM}).
In general, Eq.~(\ref{9}) 
does not satisfy Eq.~(\ref{8}). However, the ansatz (\ref{9}) was
used in 
Ref.\ \onlinecite{SYH99} only as a means to obtain the EOS from Eq.~(\ref{1}). The 
resulting EOS exhibits an excellent
agreement with simulations in 2, 3, 4, 
and 5 dimensions, provided that an
accurate $ g(\sigma )$ is used as input. 
\cite{SYH99,MV99,SYH01,GAH01,HYS02} On the other hand, if one is directly 
interested
in obtaining reliable contact values
$ g_{ij}(\sigma _{ij})$, 
then Eq.~(\ref{9}) is too crude. The simplest functional form of $G$ that 
complies with (\ref{6})--(\ref{8}) is a quadratic function of $z$: 
\begin{equation} 
G(\eta ,z)=G_{0}(\eta )+G_{1}(\eta )z+G_{2}(\eta )z^{2},  
\label{10} 
\end{equation} 
where the coefficients are explicitly given by 
\begin{subequations} 
\begin{equation}
G_{0}(\eta )=\frac{1}{1-\eta },  
\label{11a}
\end{equation}
\begin{equation}
G_{1}(\eta )= 
(2-2^{d-2}\eta )g(\sigma )-\frac{2-\eta/2}{1-\eta } , 
\label{11b} 
\end{equation}
\begin{equation}
G_{2}(\eta )=
\frac{1-\eta/2}{1-\eta }-(1-2^{d-2}\eta )g(\sigma ).  
\label{11c}
\end{equation}
\end{subequations} 
In the one-dimensional case, Eqs.~(\ref{11b}) and (\ref{11c}) lead to $G_{1}=G_{2}=0$ and we recover the exact 
result. For three-dimensional
systems, if the SPT value is used for the 
one-component contact value,
$g_{\text{SPT}}(\sigma )=(1-\eta /2+\eta 
^{2}/4)/(1-\eta )^{3}$, we
reobtain the SPT expressions
for the mixture, 
cf.\ Eq.~(\ref{15}). On the other hand, if the much
more accurate 
Carnahan-Starling \cite{CS69} (CS) expression $g_{\text{CS}}(\sigma 
)=(1-\eta /2)/(1-\eta )^{3}$ is used
as input, we arrive at the 
following
expression: 
\begin{equation} 
G(\eta ,z)=\frac{1}{1-\eta }+\frac{3}{2}\frac{
\eta (1-\eta /3)}{(1-\eta 
)^{2} }z+\frac{\eta ^{2}(1-\eta /2)}{(1-\eta
)^{3}}z^{2},\quad (d=3),  
\label{16} 
\end{equation} 
which is different from the BGHLL one and improves the latter {for $z>1$}, as comparison 
with computer simulations will show.
It should be noted that if one considers a binary mixture in the
infinite solute dilution limit, namely  $x_1 \rightarrow 0$, so that $z_{12} \rightarrow
2/(1+\sigma_2/\sigma_1)$, Eq.~(\ref{16}) yields the same result for
$g_{12}(\sigma_{12})$ as the one proposed by Matyushov  and
Ladanyi \cite{ML97} for this  quantity on the basis of exact geometrical
relations. However, the extension that the same authors propose when
there is a nonvanishing solute concentration, {i.e.} for $x_1\neq 0$
[{cf.} Eq.\ (19) in Ref.\ \onlinecite{ML97}], is different from Eq.\
(\ref{16}). We will come back to this point later when we assess the merits
of both proposals.

Of course, the quadratic form (\ref{10}) is not the only choice 
compatible with conditions (\ref{6})--(\ref{8}). Another {simple} possibility is to 
assume a rational function of the form 
\begin{equation} 
G(\eta,z)=\frac{1+\left[ A_{1}(\eta )-1\right]z}{B_{0}(\eta )+B_{1}(\eta )z 
}.  
\label{12} 
\end{equation} 
Imposing Eqs.~(\ref{6})--(\ref{8}), we get
\begin{subequations}
\begin{equation}
B_{0}(\eta )={1-\eta },
\label{13a}
\end{equation}
\begin{equation}
A_{1}(\eta )=-\frac{g(\sigma )\eta }{2}\frac{1-2^{d-1}(1-
\eta )g(\sigma )}{1-\left( 1-2^{d-1}\eta\right) g(\sigma )}, 
\label{13b}
\end{equation} 
\begin{equation}
B_{1}(\eta )=-\frac{1}{2}\frac{2-\eta -(1-\eta )\left(2-2^{d-1}\eta \right) 
g(\sigma )}{1-\left( 1-2^{d-1}\eta \right) g(\sigma)}. 
\label{13c}
\end{equation}
\label{13}
\end{subequations}
{Other functional forms for $G(\eta,z)$ complying with 
Eqs.~(\ref{6})--(\ref{8}) are possible, but the choices (\ref{10}) and 
(\ref{12}) are sufficiently representative, so we will restrict ourselves to 
them in this paper.}
\section{Comparison with simulation data\label{sec3}}
\subsection{Contact values of the radial distribution functions}
\begin{figure}[ht]
\includegraphics[width=.90\columnwidth]{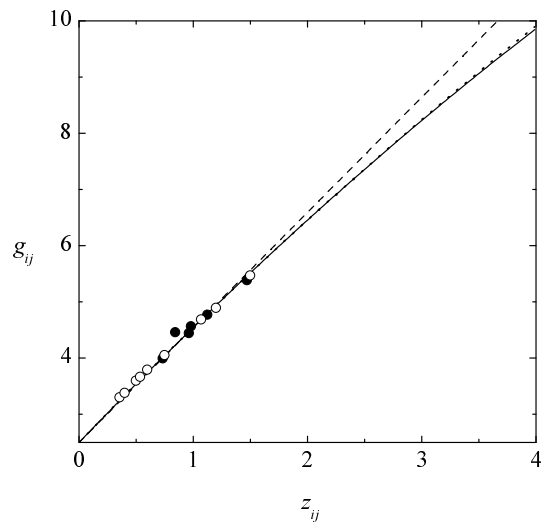}
\caption{Plot of the contact value $g_{ij}(\sigma_{ij})$ as
a function of the parameter
$z_{ij}=(\sigma _{i}\sigma_{j}/\sigma_{ij})\langle \sigma  \rangle 
/\langle\sigma ^{2}\rangle $ for hard disks ($d=2$)
at a packing fraction $\eta =0.6$.
The open circles are simulation data for three binary mixtures \protect\cite{BS01} with $\sigma_2/\sigma_1=\frac{1}{3}$ and $x_1=0.25$, 0.5, and 0.75. The filled circles  are simulation data for two binary 
mixtures\protect\cite{LS01} with $\sigma_2/\sigma_1=\frac{3}{4}$ and $x_1=0.483$ and with $\sigma_2/\sigma_1=\frac{1}{2}$ and $x_1=0.219$. The lines correspond to 
Eq.~(\protect\ref{9}) (dashed line),  
Eq.~(\protect\ref{10}) (solid line),  and Eq.~(\protect\ref{12}) (dotted 
line). \label{fig1}}
\end{figure}

\begin{figure}[ht]
\includegraphics[width=.90 \columnwidth]{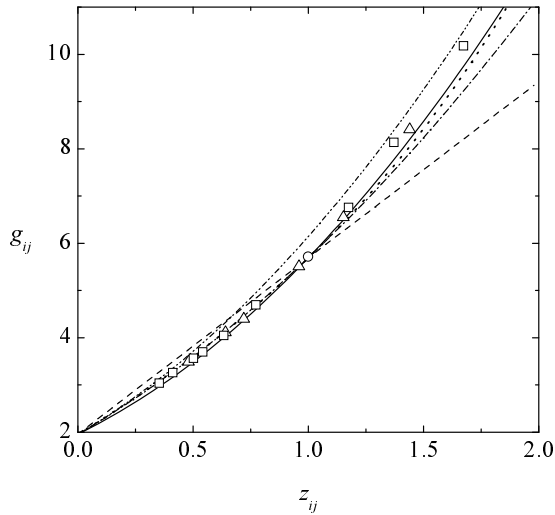}
\caption{Plot of the contact value $g_{ij}(\sigma _{ij})$ as a
function of the parameter $z_{ij}=(\sigma _{i}\sigma_{j}/\sigma_{ij})\langle 
\sigma ^{2} \rangle /\langle\sigma ^{3}\rangle $ for hard spheres ($d=3$) at 
a packing fraction $\eta =0.49$.
The symbols are simulation data for the single fluid (circle), \cite{MV99} 
three binary 
mixtures (squares) \cite{MBS97} with $\sigma_2/\sigma_1=0.3$ and $x_1=0.0625$, 0.125, and 0.25, and a ternary mixture (triangles) 
\cite{M02} with $\sigma_2/\sigma_1=\frac{2}{3}$, $\sigma_3/\sigma_1=\frac{1}{3}$ and $x_1=0.1$, $x_2=0.2$.
 The lines are, from bottom to top at the right end, 
Eq.~(\protect\ref{9}) (\mbox{-- -- --}), BGHLL (\mbox{-- $\cdot$ -- 
$\cdot$}), Eq.~(\protect\ref{12}) (\mbox{$\cdot$ $\cdot$ $\cdot$}), Eq.~(\protect\ref{10}) 
(---), and SPT (\mbox{-- $\cdot\cdot$ -- $\cdot\cdot$}). 
\label{fig2}}
\end{figure}

\begin{figure}[ht]
\includegraphics[width=.90 \columnwidth]{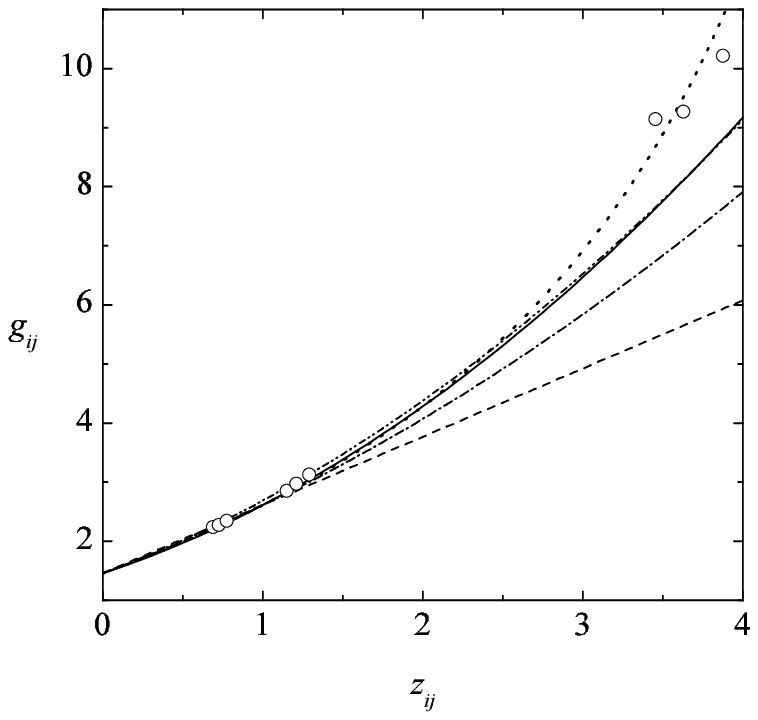}
\caption{Plot of the contact value $g_{ij}(\sigma_{ij})$ as a
function of the parameter
$z_{ij}=(\sigma _{i}\sigma_{j}/\sigma_{ij})\langle \sigma ^{2} \rangle 
/\langle\sigma ^{3}\rangle $ 
for hard spheres ($d=3$) at 
a packing fraction $\eta =0.314$.
The symbols are simulation data for  three binary 
mixtures \cite{CCHW00} with $\sigma_2/\sigma_1= \frac{1}{5}$ and $x_1=0.00311$, 0.00415, and 0.005. The lines are, from bottom to top at the right end, 
Eq.~(\protect\ref{9}) (\mbox{-- -- --}), BGHLL (\mbox{-- $\cdot$ -- 
$\cdot$}), Eq.~(\protect\ref{10}) (\mbox{---}), 
SPT (\mbox{-- $\cdot\cdot$ -- $\cdot\cdot$}), and Eq.~(\protect\ref{12}) 
(\mbox{$\cdot$ $\cdot$ $\cdot$}). \label{fig3}}
\end{figure}

\begin{figure}[ht]
\includegraphics[width=.90 \columnwidth]{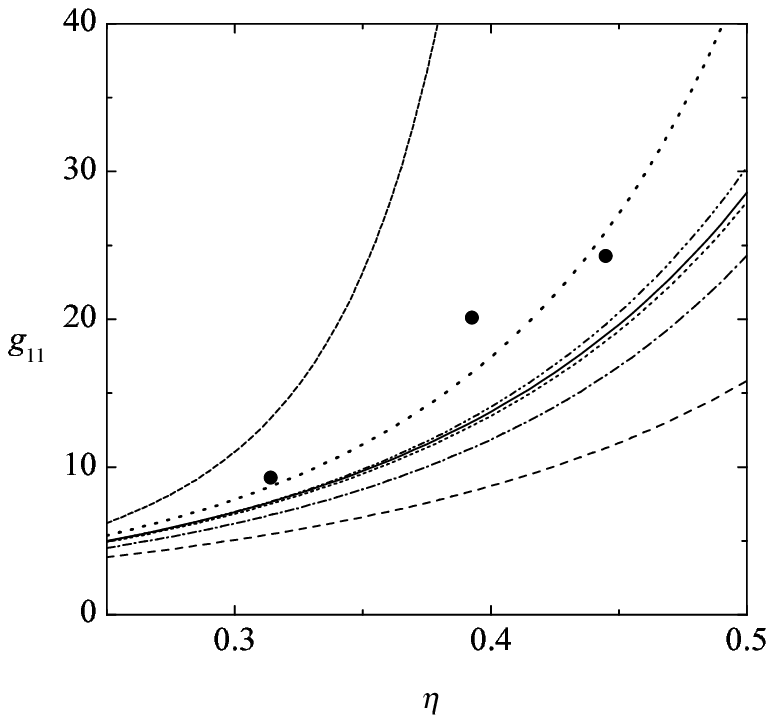}
 \caption{Plot of the contact value $g_{11}(\sigma_{1})$ as a
 function of the packing fraction $\eta$ for the 
three-dimensional binary mixture $x_1=0.005$, $\sigma_2/\sigma_1=\frac{1}{5}$ ($z_{11}=3.457$).
 The symbols are simulation data. \cite{CCHW00}  The lines are, from 
 bottom to top at the right end, 
 Eq.~(\protect\ref{9}) (\mbox{-- -- --}), BGHLL (\mbox{-- $\cdot$ -- 
$\cdot$}), Barrio--Solana 
 \cite{BS00} (\mbox{$\cdots$}), 
  Eq.~(\protect\ref{10}) (\mbox{---}), 
SPT (\mbox{-- $\cdot\cdot$ -- $\cdot\cdot$}), Eq.~(\protect\ref{12}) 
(\mbox{$\cdot$ $\cdot$ $\cdot$}), and Henderson--Chan \cite{HMLC96,YCH96,YCH97,HC98} (\mbox{- 
- -}). \label{fig4}}
 \end{figure}

\begin{figure}[ht]
\includegraphics[width=.90 \columnwidth]{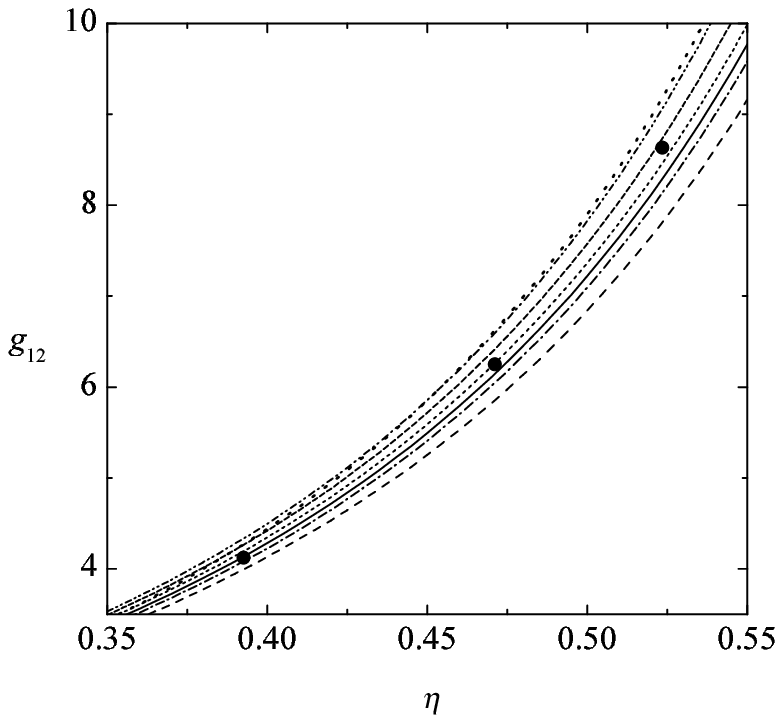}
 \caption{Plot of the contact value $g_{12}(\sigma_{12})$ as a
 function of the packing fraction $\eta$ for the 
three-dimensional binary mixture $x_1=0.00415$, $\sigma_2/\sigma_1=\frac{1}{5}$ ($z_{12}=1.210$).
 The symbols are simulation data. \cite{CCHW00}  The lines are, from 
 bottom to top at the right end, 
 Eq.~(\protect\ref{9}) (\mbox{-- -- --}), BGHLL (\mbox{-- $\cdot$ -- 
$\cdot$}),  
  Eq.~(\protect\ref{10}) (\mbox{---}), Barrio--Solana 
 \cite{BS00} (\mbox{$\cdots$}),
Henderson--Chan \cite{HMLC96,YCH96,YCH97,HC98} (\mbox{- 
- -}),
SPT (\mbox{-- $\cdot\cdot$ -- $\cdot\cdot$}),  and 
Matyushov--Ladanyi 
 \cite{ML97} (\mbox{$\cdot$ $\cdot$ $\cdot$}). \label{fig4bis}}
 \end{figure}

\begin{figure}[ht]
\includegraphics[width=.90 \columnwidth]{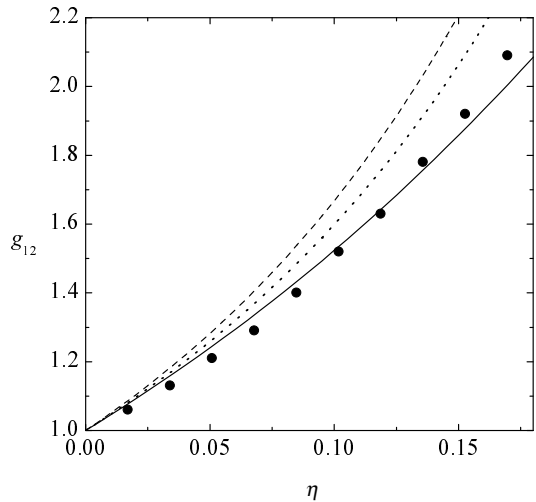}
 \caption{Plot of the contact value $g_{12}(\sigma_{12})$ as a
 function of the packing fraction $\eta$ for the 
four-dimensional binary mixture $x_1=0.5$, $\sigma_2/\sigma_1=\frac{1}{2}$ ($z_{12}=0.706$).
 The symbols are simulation data. \cite{GAH01} The lines correspond to
 Eq.~(\protect\ref{9}) (dashed line), 
  Eq.~(\protect\ref{12}) (dotted line), 
 and Eq.~(\protect\ref{10}) (solid line). 
\label{fig5}}
 \end{figure}

\begin{figure}[ht]
\includegraphics[width=.90 \columnwidth]{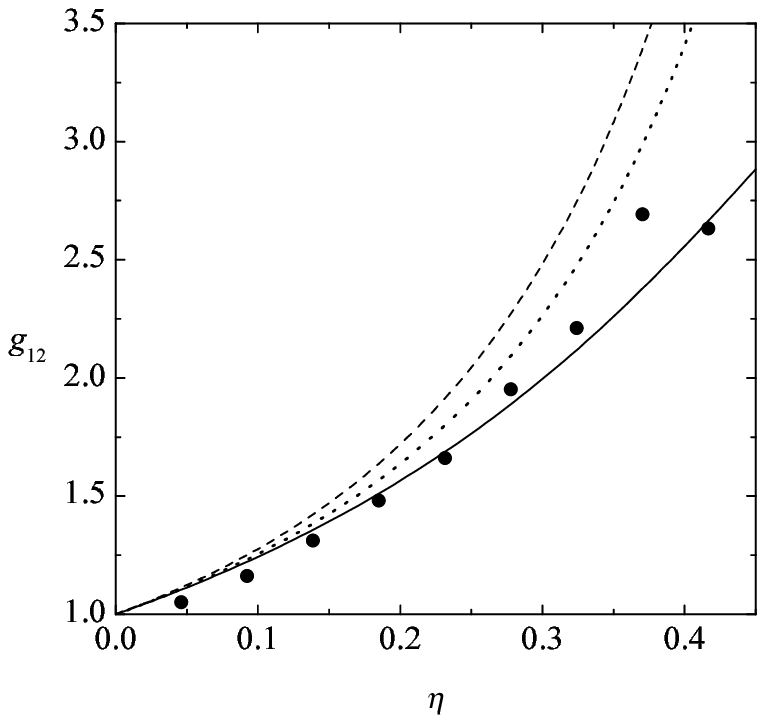}
\caption{Plot of the contact value $g_{12}(\sigma_{12})$ as a
 function of the packing fraction $\eta$ for the 
four-dimensional binary mixture $x_1=\frac{3}{4}$, $\sigma_2/\sigma_1=0.25$ ($z_{12}=0.402$).
 The symbols are simulation data. \cite{GAH01} The lines correspond to 
 Eq.~(\protect\ref{9}) (dashed line), 
  Eq.~(\protect\ref{12}) (dotted line), 
 and Eq.~(\protect\ref{10}) (solid line). 
\label{fig6}}
 \end{figure}

\begin{figure}[ht]
\includegraphics[width=.90 \columnwidth]{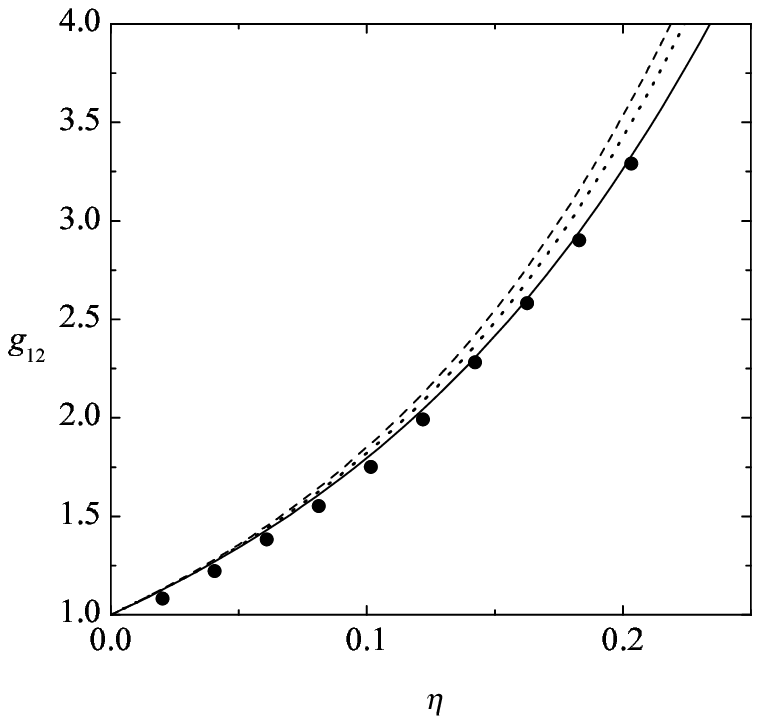}
\caption{Plot of the contact value $g_{12}(\sigma_{12})$ as a
 function of the packing fraction $\eta$ for the 
five-dimensional binary mixture $x_1=\frac{1}{2}$, $\sigma_2/\sigma_1=0.75$ ($z_{12}=0.912$).
 The symbols are simulation data. \cite{GAH01} The lines correspond to
 Eq.~(\protect\ref{9}) (dashed line), 
  Eq.~(\protect\ref{12}) (dotted line), 
 and Eq.~(\protect\ref{10}) (solid line). 
\label{fig7}}
 \end{figure}

\begin{figure}[ht]
\includegraphics[width=.90 \columnwidth]{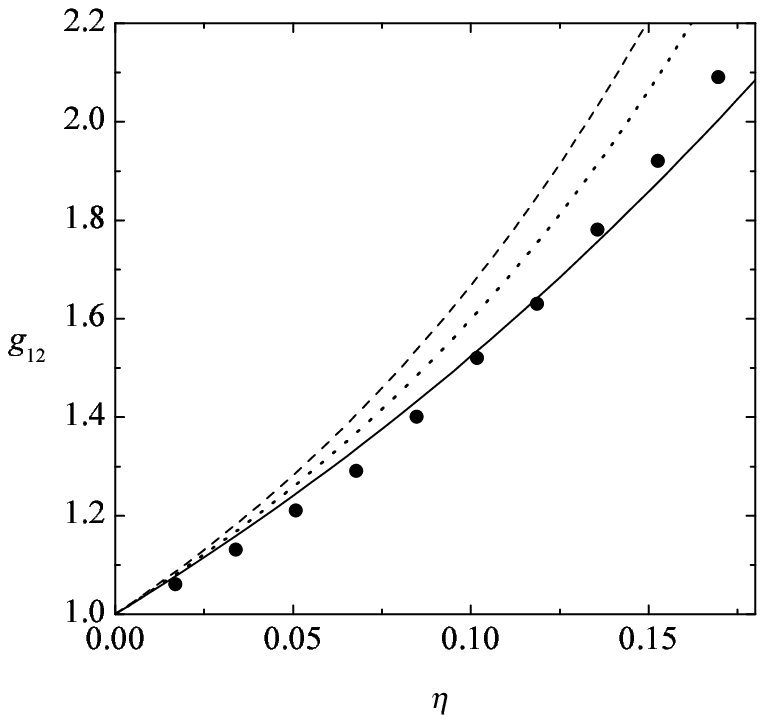}
\caption{Plot of the contact value $g_{12}(\sigma_{12})$ as a
 function of the packing fraction $\eta$ for the 
five-dimensional binary mixture $x_1=\frac{1}{2}$, $\sigma_2/\sigma_1=0.5$ ($z_{12}=0.687$).
 The symbols are simulation data. \cite{GAH01} The lines correspond to
 Eq.~(\protect\ref{9}) (dashed line), 
  Eq.~(\protect\ref{12}) (dotted line), 
 and Eq.~(\protect\ref{10}) (solid line). 
\label{fig8}}
 \end{figure}

In order to assess the utility of the new proposals for the contact 
 values of the rdf, in Figs.\ \ref{fig1}--\ref{fig8} we present results 
 for hard-core mixtures in $d=2$, 3, 4, and $5$, and the available computer 
 simulation data.
Figure \ref{fig1} shows $g_{ij}(\sigma_{ij})$ as a function of $z_{ij}$ for $d=2$ and 
$\eta=0.6$, according to the linear approximation (\ref{9}), the quadratic 
approximation (\ref{10}), and the rational approximation (\ref{12}). 
In the three cases we have used for $g(\sigma)$ the value obtained from the 
Levin[6] approximant of Erpenbeck and Luban. \cite{EL85}
The only  tabulated simulation data for $g_{ij}(\sigma_{ij})$ in the case of hard disks that we are aware of are those of Ref.~\onlinecite{BS01}.
Hence, we have included in Fig.~\ref{fig1} the simulation results for the most asymmetric mixtures considered in Ref.~\onlinecite{BS01}, namely $\sigma_2/\sigma_1=\frac{1}{3}$ with $x_1=0.25$, $0.5$, and $0.75$, and  also   simulation data extracted from Fig.\ 2  of 
Ref.\ \onlinecite{LS01}.
We observe that the quadratic and rational approximations, both consistent with condition 
(\ref{8}), are hardly distinguishable.
The three theoretical curves practically coincide in the range of values of 
$z_{ij}$ spanned by the simulations. It would be interesting to carry out 
simulations extending to the region $z_{ij}>2$ to assess the reliability of 
Eqs.~(\ref{10}) and (\ref{12}).

A comparison between theoretical predictions and simulation values 
for three-dimensional mixtures is shown in Figs.\ \ref{fig2}--\ref{fig4bis}. To 
carry out the computations in  Eqs.~(\ref{9}), (\ref{10}), and (\ref{12}), we have used the CS contact value $g_{\text{CS}}(\sigma )$. 
{Figures \ref{fig2} and \ref{fig3} show that the universality assumption (\ref{5}) is 
well supported by simulation data. Since the dependence of $g_{ij}(\sigma_{ij})$ on $z_{ij}$ is
nonlinear (note that the curvature is different from that of the 
two-dimensional case), Eq.\ (\ref{9}) only captures some kind of `average' 
behavior. Among the three quadratic functions, namely the SPT, the BGHLL, and Eq.~(\ref{10}), the best
global agreement is presented by Eq.\ (\ref{10}). The SPT prescription, Eq.\ (\ref{15}) with
$\lambda=\frac{3}{4}$, is consistent with condition (\ref{8}), but its performance is not very
good because it is pinned at a too high value at $z=1$ (one-component case). The BGHLL 
prescription, Eq.\ (\ref{15}) with $\lambda=\frac{1}{2}$, is excellent at $z=1$ (CS value), does
a very good job for $0<z<1$, but clearly underestimates the simulation data for $z>1$, as expected 
from the fact that the BGHLL is inconsistent with (\ref{8}) at $z=2$. Our recipe (\ref{10}) is
only slightly worse than the BGHLL for $z<1$ but improves it significantly 
for $z>1$. Finally, the rational function (\ref{12}) is practically 
indistinguishable from the BGHLL for $z<1$, is reasonably good for $1<z<2$, 
and is the best one in the case of the large--large rdf for
disparate mixtures, as shown by Fig.\ \ref{fig3} in the region $z\approx 4$.
Of course, none of these approximations is expected to be good enough in 
the limit of \textit{extremely}
large values of $z$, where $\ln G\sim z$.\cite{HMLC96,HBCW98,RDA01} 
The latter behavior could be incorporated by choosing an adequate functional form 
for $G(\eta,z)$
consistent with conditions (\ref{6})--(\ref{8}), but this does not 
seem to be necessary in the range 
$0\leq z \lesssim 4$.
Since  the best global agreement in the range $0\leq z\leq 2$ is
provided by the polynomial function (\ref{10}), which has a structure similar
 to the well-known BGHLL prescription in the case of $d=3$ [cf.\ 
 Eq.~(\ref{16})], we favor its use, except perhaps for very disparate 
 mixtures, where the rational approximation (\ref{12}) is preferable.}

{The refinements of the BGHLL expressions recently proposed by Henderson and Chan \cite{HMLC96,YCH96,YCH97,HC98}
and by Barrio and Solana \cite{BS00} are not shown in Figs.\ \ref{fig2} and \ref{fig3} because
they do not belong to the class of approximations satisfying the universality assumption (\ref{5}).
In addition, they are restricted to the case of \textit{binary} mixtures. 
Both approximations differ
in practice from the BGHLL only in $g_{11}(\sigma_1)$, where species 1 refer to the big spheres 
($\sigma_1>\sigma_2$). Figure \ref{fig4} shows $g_{11}(\sigma_1)$ versus $\eta$ for the 
three-dimensional binary mixture $x_1=0.005$, $\sigma_2/\sigma_1=\frac{1}{5}$ (which 
corresponds to $z_{11}=3.457$). The figure confirms that the best agreement 
is obtained with the rational
approximation (\ref{12}). Henderson and Chan's approximation, which incorporates the exact 
asymptotic behavior $\ln g_{11}(\sigma_1)\sim \sigma_1$, gives too 
high values. Barrio and Solana's expression improves the BGHLL value, but is 
slightly worse than the quadratic approximation (\ref{10}).}

Figure \ref{fig4bis} presents a plot of $g_{12}(\sigma_{12})$ as a function of $\eta$ for
the three-dimensional binary mixture characterized by $x_1=0.00415$ and
$\sigma_2/\sigma_1=\frac{1}{5}$ (which corresponds to a value of
$z_{12}=1.210$) as given by different approximations. The rational
approximation given by Eq.\ (\ref{12}) has not been included in the figure
to avoid overcrowding of the curves, but it is practically
indistinguishable from the BGHLL approximation in this case. Clearly the
best agreement between theory and simulation is provided by the
approximations of Barrio and Solana, \cite{BS00} of Henderson and Chan,
\cite{HC98} and by our Eq.\ (\ref{10}), which are all of comparable
accuracy and certainly superior to the approximation proposed  in Eq. (19)
of the paper by Matyushov and Ladanyi. \cite{ML97}

{Of course, the most physically relevant situations correspond to 
 hard spheres ($d=3$) and, to
a lesser extent, disks ($d=2$). On the other hand, it seems desirable that a proposal for 
$g_{ij}(\sigma_{ij})$ be valid for any dimensionality $d$. Moreover, a number of 
recent papers deal with systems of hard hyperspheres. 
\cite{SM98,FP99,BMC99,PS00,YSH00,GAH01,FSL02,EAGB02}
Figures \ref{fig5}--\ref{fig8} show $g_{12}(\sigma_{12})$ versus $\eta$ for binary mixtures in $d=4$ and $d=5$. The contact values of the one-component
system that we have used in the computations have been obtained 
from the EOS derived for these systems by Luban and Michels.\cite{LM90} The values of the parameter $z_{12}$ are in the range $0.4<z_{12}<1$
for the cases considered in Figs.\ \ref{fig5}--\ref{fig8}. It is observed that 
in this range the quadratic approximation (\ref{10}) exhibits an excellent agreement 
with the simulation data.}

\subsection{Equation of state}

Having examined the accuracy of the proposed contact values, we will 
now consider their performance in terms of the compressibility factor they 
lead to. In this regard, Eq.~(\ref{10}) has the advantage over 
Eq.~(\ref{12}) that, when inserted into Eq.~(\ref{1}), one gets a closed 
expression for the compressibility factor, in terms of the packing fraction 
$\eta $ and the first few moments $\langle \sigma ^{n}\rangle$, $n\leq d$. 
This expression is meaningful even for polydisperse mixtures. The result is 
\begin{eqnarray} 
Z_{\text{m}}(\eta) &=&1+2^{d-2}\frac{\eta }{1-\eta }\left[ 
2(S_{0}-2S_{1}+S_{2})+(S_{1}-S_{2})\eta \right]  \nonumber \\ 
&&+\left[Z_{\text{s}}(\eta )-1\right] \left[ 
2S_{1}-S_{2}+2^{d-2}(S_{2}-S_{1})\eta \right] ,  
\label{14} 
\end{eqnarray} 
where $Z_{\text{s}}(\eta )=1+2^{d-1}\eta g(\sigma )$ is the compressibility 
factor of the one-component system and the coefficients $S_{m}$ are given by
\begin{equation}
S_{m}=2^{-(d-m)}\frac{\langle\sigma^{d-1}\rangle^{m}}{\langle 
\sigma^{d}\rangle^{m+1}}\sum_{n=0}^{d-m}\binom{d-m}{n}
{\langle \sigma^{n+m}\rangle }{\langle \sigma ^{d-n}\rangle}.
\end{equation}

In the two-dimensional case, Eq.~(\ref{14}) becomes 
\begin{equation} 
Z_{\text{m}}(\eta )=\frac{1}{1-\eta }+\frac{\langle \sigma \rangle ^{2}}{ 
\langle \sigma ^{2}\rangle }\left[ Z_{\text{s}}(\eta)-\frac{1}{1-\eta } 
\right] ,\quad (d=2). 
\label{new2}
\end{equation} 
It is worth noticing that this EOS coincides with the one obtained from 
Eq.~(\ref{9}) for $d=2$. 
This illustrates the fact that two different 
proposals for the contact values $g_{ij}(\sigma _{ij})$ can yield the same 
EOS when inserted into Eq.~(\ref{1}). 
{Let us analyze this point with more detail. Subtracting Eqs.~(\ref{9}) and (\ref{10}), one has 
\beq
\Delta G(\eta,z)=
\left[\frac{1-\eta/2}{1-\eta }
-(1-2^{d-2}\eta )g(\sigma )\right]z(1-z),
\label{new1}
\eeq
where $\Delta G(\eta,z)$ denotes the difference between the linear and the quadratic approximations.
Thus, the compressibility factors  obtained from the linear and quadratic forms for $G(\eta,z)$ only differ by a term proportional to $\sum_{i,j} x_i x_j\sigma_{ij}^d z_{ij}(1-z_{ij})$. It turns out that this term vanishes in the two-dimensional case, so the linear and quadratic approximations yield the same EOS (\ref{new2}).
This fact}  also shows that a rather 
crude approximation such as Eq.~(\ref{9}) may lead to an extremely good EOS. 
\cite{SYH99,SYH01,GAH01,HYS02} 

For three-dimensional mixtures, 
Eq.~(\ref{14}) becomes 
\begin{eqnarray}
 Z_{\text{m}}(\eta )&=&\frac{1}{1-\eta }+\frac{\langle \sigma
\rangle \langle 
 \sigma ^{2}\rangle }{\langle \sigma^{3}\rangle }\left( 1-\eta +\frac{ 
 \langle \sigma ^{2}\rangle ^{2}}{\langle\sigma \rangle \langle \sigma 
 ^{3}\rangle }\eta \right) \nonumber\\
&&\times\left[ Z_{\text{s}}(\eta )-\frac{1}{1-\eta 
 }\right] ,\quad (d=3). 
 \end{eqnarray} 
In particular, when the CS EOS $Z_{\text{s}}(\eta )=(1+\eta +\eta ^{2}-\eta^{3})/(1-\eta )^{3}$ is used as input, we get {the following extended CS EOS:}
\begin{equation}
Z_{\text{eCS-II}}(\eta )=Z_{\text{BMCSL}}(\eta)-\frac{\eta 
^{3}\langle \sigma ^{2}\rangle}{(1-\eta)^{2}\langle\sigma
^{3}\rangle ^{2}}\left( \langle \sigma \rangle 
\langle\sigma ^{3}\rangle-\langle \sigma ^{2}\rangle ^{2}\right) 
,
\label{eCS2}
\end{equation} 
where the compressibility factor associated with the BMCSL EOS\cite{B70,MCSL71} 
is given in the present notation by 
\begin{equation}
Z_{\text{BMCSL}}(\eta )=\frac{1}{1-\eta }+\frac{3\eta \langle \sigma\rangle 
\langle\sigma ^{2}\rangle }{(1-\eta )^{2}\langle \sigma 
^{3}\rangle}+\frac{\eta^{2}(3-\eta )\langle \sigma ^{2}\rangle^{3}}{(1-\eta 
)^{3}\langle \sigma ^{3}\rangle ^{2}}.
\end{equation}
Note that Eq.~(\ref{eCS2}) is different from the extended CS EOS 
obtained from Eq.~(\ref{9}), namely \cite{SYH99}
\begin{equation}
Z_{\text{eCS-I}}(\eta )=Z_{\text{BMCSL}}(\eta 
)+\frac{\eta 
^{3}\langle \sigma ^{2}\rangle}{(1-\eta)^{3}\langle\sigma
^{3}\rangle ^{2}}
\left( \langle \sigma \rangle 
\langle\sigma ^{3}\rangle-\langle \sigma ^{2}\rangle ^{2}\right) 
.
\label{eCS1}
\end{equation}

\begin{figure}[ht]
\includegraphics[width=.90 \columnwidth]{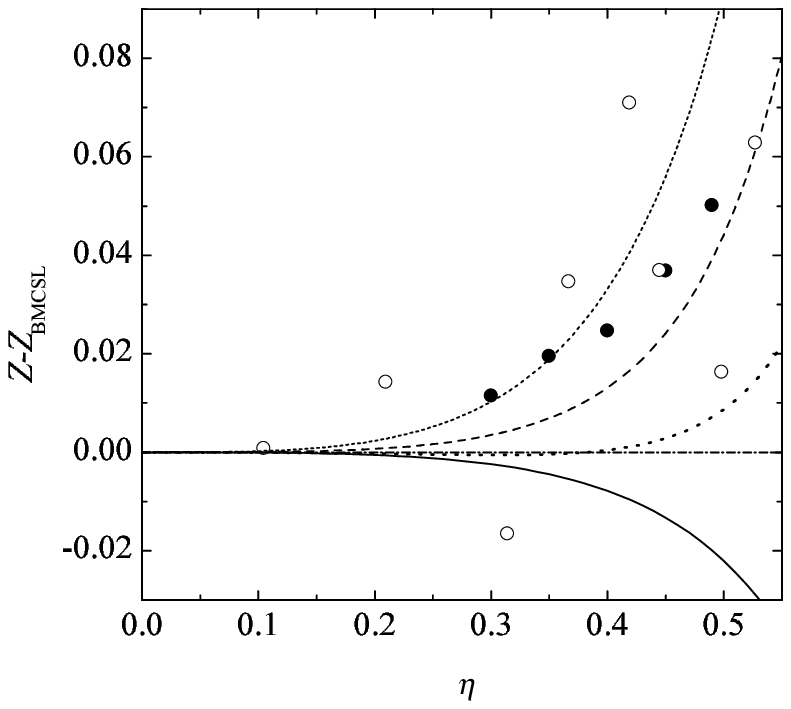}
 \caption{Deviation of the compressibility factor from the BMCSL value, 
  as a
 function of the packing fraction $\eta$ for an equimolar
three-dimensional binary mixture with $\sigma_2/\sigma_1=0.6$.
 The open \cite{YCH96} and filled \cite{BMLS96} circles are simulation data. 
 The lines are, from bottom to top at the right end, the eCS-II EOS 
 (\protect\ref{eCS2}) (\mbox{---}), the EOS obtained from the rational approximation (\protect\ref{12}) (\mbox{$\cdot$ $\cdot$ $\cdot$}), the eCS-I EOS  (\protect\ref{eCS1}) (\mbox{-- -- --}),
 and  Barrio and Solana's EOS (\mbox{$\cdots$}). \protect\cite{BS00}
\label{fig9}}
 \end{figure}

\begin{figure}[ht]
\includegraphics[width=.90 \columnwidth]{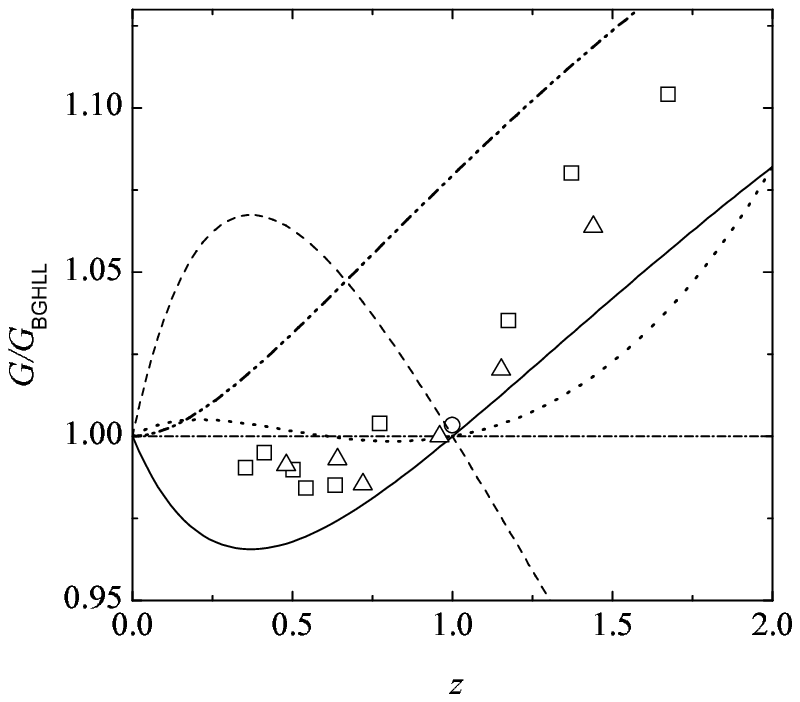}
 \caption{Plot of the ratio $G(\eta ,z)/G_{\text{BGHLL}}(\eta ,z)$ as a
 function of the parameter
 $z$ for hard spheres ($d=3$) at 
 a packing fraction $\eta =0.49$.
The symbols are simulation data for the single fluid (circle), \cite{MV99} 
three binary 
mixtures (squares) \cite{MBS97} with $\sigma_2/\sigma_1=0.3$ and $x_1=0.0625$, 0.125, and 0.25, and a ternary mixture (triangles) 
\cite{M02} with $\sigma_2/\sigma_1=\frac{2}{3}$, $\sigma_3/\sigma_1=\frac{1}{3}$ and $x_1=0.1$, $x_2=0.2$.
 The lines correspond to Eq.~(\protect\ref{9}) (\mbox{-- -- --}), Eq.~(\protect\ref{10}) (\mbox{---}), Eq.~(\protect\ref{12}) (\mbox{$\cdot$ $\cdot$ $\cdot$}), and SPT (\mbox{-- $\cdot\cdot$ -- $\cdot\cdot$}). 
 \label{fig10}}
\end{figure}

Since simulation data indicate that the BMCSL EOS tends to underestimate
the 
compressibility factor, it is obvious that
the performance of $Z_{\text{eCS-I}}$ is, {paradoxically}, better than that of $Z_{\text{eCS-II}}$. This is explicitly 
shown in Fig.~\ref{fig9} where, for comparison, we have also included the 
theoretical results that follow from the recent (very accurate) proposal by 
Barrio and Solana \cite{BS00} for the EOS. 
So once again we find that  better contact values do not necessarily lead to 
better compressibility factors, {as already seen in the 
two-dimensional case}. One plausible explanation {for the better performance of $Z_{\text{eCS-I}}$ with respect to $Z_{\text{eCS-II}}$} might reside on 
the use of the CS EOS $Z_{\text{s}}(\eta )$, but we have checked that if the 
Pad\'{e} [4,3] of Sanchez \cite{S94} or the very accurate EOS of 
Malijevsk\'{y} and Veverka \cite{MV99} are used instead, the results do not 
change. Thus the source of this effect is a 
{fortunate} compensation of errors {in the linear approximation (\ref{9}) related to the fact that the compressibility factor involves a  weighted average of the individual contact values $g_{ij}(\sigma_{ij})$  [cf.\ Eq.\ (\ref{1})]}. This argument is suggested by the following 
observation. In Fig.~\ref{fig10} we present a plot of the ratio $G(\eta ,z)/G_{\text{BGHLL}}(\eta ,z)$ as a function of $z$ for $\eta=0.49$ with $G(\eta ,z)$ 
given by Eqs.\ (\ref{9}) and (\ref{10}). {For completeness, also the ratios corresponding to the SPT (\ref{15}) and the rational approximation (\ref{12}) are plotted}. Properly 
reduced simulation results have also been included. From this plot it is 
fair to conclude that the approximation given by Eq.~(\ref{10}) is globally more 
accurate than those obtained with either Eq.~(\ref{9}) or with $G_{\text{BGHLL}}(\eta ,z)$ and that the BGHLL contact values are 
better than the linear approximation. However, the quadratic approximation 
\textit{always} underestimates the simulation results (and therefore it is 
only natural that it will always produce an underestimation of the 
compressibility factor), while both the linear approximation and the BGHLL 
approximation overestimate the simulation results if $z_{ij}\leq 1$ but 
underestimate them if $ z_{ij}\geq 1$. The net result is 
that $Z_{\text{eCS-II}}(\eta )$ is in poorer agreement with the 
simulation results for the compressibility factor than either 
$Z_{\text{eCS-I}}(\eta )$ or $Z_{\text{BMCSL}}(\eta )$, the extended CS EOS 
obtained from Eq.~(\ref{9}) providing the best overall agreement.
Nevertheless, an important asset of $Z_{\text{eCS-II}}$, not shared by either
$Z_{\text{eCS-I}}$ or $Z_{\text{BMCSL}}$, is that it predicts
demixing. This result provides further support to the analysis performed
by Regnaut, Dyan, and Amokrane \cite{RDA01} in which the verification of
condition (\ref{4}) is of key importance for the existence of
demixing. We will address this issue in more detail elsewhere.

For $d=4$ and $d=5$ (not shown), {however, the compressibility factors derived
from the linear approximation given by Eq.\ (\ref{9}) \cite{SYH99} turn out 
 to be slightly less accurate   than those obtained 
from the use of  either Eq.\ (\ref{10}) or Eq.~(\ref{12}). Thus, for these high dimensionalities, the expectation that a better overall behavior of the contact values correlates with   a better performance of the associated EOS actually holds true.}

\section{Conclusion\label{sec4}}

In summary, in this {paper} we have introduced {a universality assumption, Eq.\ (\ref{5}), for the contact values of the rdf of a hard-sphere mixture with  arbitrary
number of components and arbitrary dimensionality. Three known consistency conditions, Eqs.\ 
(\ref{2})--(\ref{4}), allow us to fix the values of the universal function $G(\eta,z)$ at $z=0$
[cf.\ Eq.\ (\ref{6})], $z=1$ [cf.\ Eq.\ (\ref{7})], and $z=2$ [cf.\ Eq.\ (\ref{8})], the two latter
in terms of the  contact value of the one-component rdf. This implies that any reasonable 
three-parameter form of $G(\eta,z)$ as a function of $z$ can provide a very good approximate
representation of $g_{ij}(\sigma_{ij})$ regardless of the number of components, provided a
good EOS for the single fluid system is used. We have illustrated this possibility with 
two specific proposals: a quadratic function, Eq.~(\ref{10}), and a rational function, Eq.\
(\ref{12}).} In $d=1$, they reduce to the 
exact result, while for $d=3$ they represent an improvement over the BGHLL 
values, as well as over those of their refinements,  in the 
quantitative
agreement with the simulation results. For $d=2$, 4, and $5$, 
they compare
rather well with the (few) available simulation
results. Their 
potential use in connection with the generation of the
entire rdf 
$g_{ij}(r)$ for three-dimensional
mixtures within the Rational Function 
Approximation method is currently
under investigation.

The  relationship between the thermodynamic properties and the contact
values of the rdf in hard-core fluids is in principle so straightforward
that the importance of explicit and accurate approximations for the latter
can be hardly overemphasized.  In one-component hard-core systems it is
certainly true that a more accurate contact value of the rdf leads
directly to a better EOS. On the other hand, our results in this paper
indicate that, due to the fact that for mixtures the EOS (or equivalently
the compressibility factor) involves a summation over species
indices in which the different contact values are included, one does not
{\em always\/} obtain a more accurate EOS from seemingly better
approximations to $g_{ij}(\sigma_{ij})$. In fact, as exemplified in the
case of hard-disk mixtures ($d=2$), it is possible to obtain exactly the
same EOS with two different approximations for the contact
values of the rdf. Further, the poorer agreement of $Z_{\text{eCS-II}}$ with
simulation data than either $Z_{\text{eCS-I}}$ or $Z_{\text{BMCSL}}$ mentioned in the
previous Section is also a reflection of the above assertion, the reason resting on a `fortunate' compensation of
errors. In any event, in some specific applications (e.g.\ the Enskog
kinetic theory) it is only the contact values of the rdf that are
required. In this respect, it is fair to conclude that our two new
proposals provide in general a reasonably accurate approximation to 
$g_{ij}(\sigma_{ij})$ (as compared to the available simulation data) for a
hard-core mixture with an arbitrary number of components and arbitrary
dimensionality. Of course the scarcity of simulation results for these
systems precludes a more definite conclusion. In any case, we hope that
the availability  of the new (explicit) expressions for the contact values
of the rdf of hard-core mixtures in any dimensionality may serve as a
further motivation to carry out yet more simulations of these systems.

\acknowledgments
The authors are grateful to Alexander Malijevsk\'y for providing simulations data of ternary hard-sphere mixtures prior to publication.
A.S. and S.B.Y. acknowledge partial support from the Ministerio de Ciencia y Tecnolog\'{\i}a
 (Spain) and FEDER through grant No.\ BFM2001-0718


\begin{thebibliography}{41}
\expandafter\ifx\csname natexlab\endcsname\relax\def\natexlab#1{#1}\fi
\expandafter\ifx\csname bibnamefont\endcsname\relax
  \def\bibnamefont#1{#1}\fi
\expandafter\ifx\csname bibfnamefont\endcsname\relax
  \def\bibfnamefont#1{#1}\fi
\expandafter\ifx\csname citenamefont\endcsname\relax
  \def\citenamefont#1{#1}\fi
\expandafter\ifx\csname url\endcsname\relax
  \def\url#1{\texttt{#1}}\fi
\expandafter\ifx\csname urlprefix\endcsname\relax\def\urlprefix{URL }\fi
\providecommand{\bibinfo}[2]{#2}
\providecommand{\eprint}[2][]{\url{#2}}

\bibitem[{\citenamefont{Lebowitz}(1964)}]{L64}
\bibinfo{author}{\bibfnamefont{J.~L.} \bibnamefont{Lebowitz}},
  \bibinfo{journal}{Phys. Rev. A} \textbf{\bibinfo{volume}{133}},
  \bibinfo{pages}{895} (\bibinfo{year}{1964}).

\bibitem[{\citenamefont{Boubl\'{\i}k}(1970)}]{B70}
\bibinfo{author}{\bibfnamefont{T.}~\bibnamefont{Boubl\'{\i}k}},
  \bibinfo{journal}{J. Chem. Phys.} \textbf{\bibinfo{volume}{53}},
  \bibinfo{pages}{471} (\bibinfo{year}{1970}).

\bibitem[{\citenamefont{Mansoori et~al.}(1971)\citenamefont{Mansoori, Carnahan,
  Starling, and T.~W.~Leland}}]{MCSL71}
\bibinfo{author}{\bibfnamefont{G.~A.} \bibnamefont{Mansoori}},
  \bibinfo{author}{\bibfnamefont{N.~F.} \bibnamefont{Carnahan}},
  \bibinfo{author}{\bibfnamefont{K.~E.} \bibnamefont{Starling}},
  \bibnamefont{and}
  \bibinfo{author}{\bibfnamefont{J.}~\bibnamefont{T.~W.~Leland}},
  \bibinfo{journal}{J. Chem. Phys.} \textbf{\bibinfo{volume}{54}},
  \bibinfo{pages}{1523} (\bibinfo{year}{1971}).

\bibitem[{\citenamefont{Lebowitz et~al.}(1965)\citenamefont{Lebowitz, Helfand,
  and Praestgaard}}]{LHP65}
\bibinfo{author}{\bibfnamefont{J.~L.} \bibnamefont{Lebowitz}},
  \bibinfo{author}{\bibfnamefont{E.}~\bibnamefont{Helfand}}, \bibnamefont{and}
  \bibinfo{author}{\bibfnamefont{E.}~\bibnamefont{Praestgaard}},
  \bibinfo{journal}{J. Chem. Phys.} \textbf{\bibinfo{volume}{43}},
  \bibinfo{pages}{774} (\bibinfo{year}{1965}).

\bibitem[{\citenamefont{Rosenfeld}(1988)}]{R88}
\bibinfo{author}{\bibfnamefont{Y.}~\bibnamefont{Rosenfeld}},
  \bibinfo{journal}{J. Chem. Phys.} \textbf{\bibinfo{volume}{89}},
  \bibinfo{pages}{4272} (\bibinfo{year}{1988}).

\bibitem[{\citenamefont{Grundke and Henderson}(1972)}]{GH72}
\bibinfo{author}{\bibfnamefont{E.~W.} \bibnamefont{Grundke}} \bibnamefont{and}
  \bibinfo{author}{\bibfnamefont{D.}~\bibnamefont{Henderson}},
  \bibinfo{journal}{Mol. Phys.} \textbf{\bibinfo{volume}{24}},
  \bibinfo{pages}{269} (\bibinfo{year}{1972}).

\bibitem[{\citenamefont{Lee and Levesque}(1973)}]{LL73}
\bibinfo{author}{\bibfnamefont{L.~L.} \bibnamefont{Lee}} \bibnamefont{and}
  \bibinfo{author}{\bibfnamefont{D.}~\bibnamefont{Levesque}},
  \bibinfo{journal}{Mol. Phys.} \textbf{\bibinfo{volume}{26}},
  \bibinfo{pages}{1351} (\bibinfo{year}{1973}).

\bibitem[{\citenamefont{Henderson et~al.}(1996)\citenamefont{Henderson,
  Malijevsk\'{y}, Lab\'{\i}k, and Chan}}]{HMLC96}
\bibinfo{author}{\bibfnamefont{D.}~\bibnamefont{Henderson}},
  \bibinfo{author}{\bibfnamefont{A.}~\bibnamefont{Malijevsk\'{y}}},
  \bibinfo{author}{\bibfnamefont{S.}~\bibnamefont{Lab\'{\i}k}},
  \bibnamefont{and} \bibinfo{author}{\bibfnamefont{K.~Y.} \bibnamefont{Chan}},
  \bibinfo{journal}{Mol. Phys.} \textbf{\bibinfo{volume}{87}},
  \bibinfo{pages}{273} (\bibinfo{year}{1996}).

\bibitem[{\citenamefont{Yau et~al.}(1996)\citenamefont{Yau, Chan, and
  Henderson}}]{YCH96}
\bibinfo{author}{\bibfnamefont{D.~H.~L.} \bibnamefont{Yau}},
  \bibinfo{author}{\bibfnamefont{K.-Y.} \bibnamefont{Chan}}, \bibnamefont{and}
  \bibinfo{author}{\bibfnamefont{D.}~\bibnamefont{Henderson}},
  \bibinfo{journal}{Mol. Phys.} \textbf{\bibinfo{volume}{88}},
  \bibinfo{pages}{1237} (\bibinfo{year}{1996}).

\bibitem[{\citenamefont{Yau et~al.}(1997)\citenamefont{Yau, Chan, and
  Henderson}}]{YCH97}
\bibinfo{author}{\bibfnamefont{D.~H.~L.} \bibnamefont{Yau}},
  \bibinfo{author}{\bibfnamefont{K.-Y.} \bibnamefont{Chan}}, \bibnamefont{and}
  \bibinfo{author}{\bibfnamefont{D.}~\bibnamefont{Henderson}},
  \bibinfo{journal}{Mol. Phys.} \textbf{\bibinfo{volume}{91}},
  \bibinfo{pages}{1813} (\bibinfo{year}{1997}).

\bibitem[{\citenamefont{Henderson and Chan}(1998)}]{HC98}
\bibinfo{author}{\bibfnamefont{D.}~\bibnamefont{Henderson}} \bibnamefont{and}
  \bibinfo{author}{\bibfnamefont{K.~Y.} \bibnamefont{Chan}},
  \bibinfo{journal}{J. Chem. Phys.} \textbf{\bibinfo{volume}{108}},
  \bibinfo{pages}{9946} (\bibinfo{year}{1998}).

\bibitem{MHC99}
D. Matyushov, D. Henderson, and K.-Y. Chan, Mol. Phys. \textbf{96}, 1813 (1999).

\bibitem{ML97}
D. V. Matyushov and B. M. Ladanyi, J. Chem. Phys. \textbf{107}, 5815 (1997)

\bibitem[{\citenamefont{Barrio and Solana}(2000)}]{BS00}
\bibinfo{author}{\bibfnamefont{C.}~\bibnamefont{Barrio}} \bibnamefont{and}
  \bibinfo{author}{\bibfnamefont{J.~R.} \bibnamefont{Solana}},
  \bibinfo{journal}{J. Chem. Phys.} \textbf{\bibinfo{volume}{113}},
  \bibinfo{pages}{10180} (\bibinfo{year}{2000}).

\bibitem[{\citenamefont{Jenkins and Mancini}(1987)}]{JM87}
\bibinfo{author}{\bibfnamefont{J.~T.} \bibnamefont{Jenkins}} \bibnamefont{and}
  \bibinfo{author}{\bibfnamefont{F.}~\bibnamefont{Mancini}},
  \bibinfo{journal}{J. Appl. Mech.} \textbf{\bibinfo{volume}{54}},
  \bibinfo{pages}{27} (\bibinfo{year}{1987}).

\bibitem[{\citenamefont{Santos et~al.}(1999)\citenamefont{Santos, Yuste, and
  {L\'opez de Haro}}}]{SYH99}
\bibinfo{author}{\bibfnamefont{A.}~\bibnamefont{Santos}},
  \bibinfo{author}{\bibfnamefont{S.~B.} \bibnamefont{Yuste}}, \bibnamefont{and}
  \bibinfo{author}{\bibfnamefont{M.}~\bibnamefont{{L\'opez de Haro}}},
  \bibinfo{journal}{Mol. Phys.} \textbf{\bibinfo{volume}{96}},
  \bibinfo{pages}{1} (\bibinfo{year}{1999}).

\bibitem[{\citenamefont{{L\'opez de Haro} et~al.}(2002)\citenamefont{{L\'opez
  de Haro}, Yuste, and Santos}}]{HYS02}
\bibinfo{author}{\bibfnamefont{M.}~\bibnamefont{{L\'opez de Haro}}},
  \bibinfo{author}{\bibfnamefont{S.~B.} \bibnamefont{Yuste}}, \bibnamefont{and}
  \bibinfo{author}{\bibfnamefont{A.}~\bibnamefont{Santos}},
  \emph{\bibinfo{title}{{The equation of state of additive hard-disk fluid
  mixtures: A critical analysis of two recent proposals}}},
  \bibinfo{howpublished}{submitted to Phys. Rev. E} (\bibinfo{year}{2002}).

\bibitem[{\citenamefont{Malijevsk\'{y} and Veverka}(1999)}]{MV99}
\bibinfo{author}{\bibfnamefont{A.}~\bibnamefont{Malijevsk\'{y}}}
  \bibnamefont{and} \bibinfo{author}{\bibfnamefont{J.}~\bibnamefont{Veverka}},
  \bibinfo{journal}{Phys. Chem. Chem. Phys.} \textbf{\bibinfo{volume}{1}},
  \bibinfo{pages}{4267} (\bibinfo{year}{1999}).

\bibitem[{\citenamefont{Cao et~al.}(2000)\citenamefont{Cao, Chan, Henderson,
  and Wang}}]{CCHW00}
\bibinfo{author}{\bibfnamefont{D.}~\bibnamefont{Cao}},
  \bibinfo{author}{\bibfnamefont{K.-Y.} \bibnamefont{Chan}},
  \bibinfo{author}{\bibfnamefont{D.}~\bibnamefont{Henderson}},
  \bibnamefont{and} \bibinfo{author}{\bibfnamefont{W.}~\bibnamefont{Wang}},
  \bibinfo{journal}{Mol. Phys.} \textbf{\bibinfo{volume}{98}},
  \bibinfo{pages}{619} (\bibinfo{year}{2000}).

\bibitem[{\citenamefont{Gonz\'alez-Melchor
  et~al.}(2001)\citenamefont{Gonz\'alez-Melchor, Alejandre, and {L\'opez de
  Haro}}}]{GAH01}
\bibinfo{author}{\bibfnamefont{M.}~\bibnamefont{Gonz\'alez-Melchor}},
  \bibinfo{author}{\bibfnamefont{J.}~\bibnamefont{Alejandre}},
  \bibnamefont{and} \bibinfo{author}{\bibfnamefont{M.}~\bibnamefont{{L\'opez de
  Haro}}}, \bibinfo{journal}{J. Chem. Phys.} \textbf{\bibinfo{volume}{114}},
  \bibinfo{pages}{4905} (\bibinfo{year}{2001}).

\bibitem[{\citenamefont{Giunta et~al.}(1985)\citenamefont{Giunta, Abramo, and
  Caccamo}}]{GAC85}
\bibinfo{author}{\bibfnamefont{G.}~\bibnamefont{Giunta}},
  \bibinfo{author}{\bibfnamefont{M.~C.} \bibnamefont{Abramo}},
  \bibnamefont{and} \bibinfo{author}{\bibfnamefont{C.}~\bibnamefont{Caccamo}},
  \bibinfo{journal}{Mol. Phys.} \textbf{\bibinfo{volume}{56}},
  \bibinfo{pages}{319} (\bibinfo{year}{1985}).

\bibitem[{\citenamefont{Yuste et~al.}(1998)\citenamefont{Yuste, Santos, and
  {L\'opez de Haro}}}]{YSH98}
\bibinfo{author}{\bibfnamefont{S.~B.} \bibnamefont{Yuste}},
  \bibinfo{author}{\bibfnamefont{A.}~\bibnamefont{Santos}}, \bibnamefont{and}
  \bibinfo{author}{\bibfnamefont{M.}~\bibnamefont{{L\'opez de Haro}}},
  \bibinfo{journal}{J. Chem. Phys.} \textbf{\bibinfo{volume}{108}},
  \bibinfo{pages}{3683} (\bibinfo{year}{1998}).

\bibitem[{\citenamefont{Garz\'o and Dufty}(1999)}]{GD99}
\bibinfo{author}{\bibfnamefont{V.}~\bibnamefont{Garz\'o}} \bibnamefont{and}
  \bibinfo{author}{\bibfnamefont{J.~W.} \bibnamefont{Dufty}},
  \bibinfo{journal}{Phys. Rev. E} \textbf{\bibinfo{volume}{60}},
  \bibinfo{pages}{5706} (\bibinfo{year}{1999}).

\bibitem[{\citenamefont{Hamad}(1994)}]{H94}
\bibinfo{author}{\bibfnamefont{E.}~\bibnamefont{Hamad}}, \bibinfo{journal}{J.
  Chem. Phys.} \textbf{\bibinfo{volume}{101}}, \bibinfo{pages}{10195}
  (\bibinfo{year}{1994}).


\bibitem[{\citenamefont{Henderson et~al.}(1998)\citenamefont{Henderson, Boda,
  Chan, and Wasan}}]{HBCW98}
\bibinfo{author}{\bibfnamefont{D.}~\bibnamefont{Henderson}},
  \bibinfo{author}{\bibfnamefont{D.}~\bibnamefont{Boda}},
  \bibinfo{author}{\bibfnamefont{K.~Y.} \bibnamefont{Chan}}, \bibnamefont{and}
  \bibinfo{author}{\bibfnamefont{D.~T.} \bibnamefont{Wasan}},
  \bibinfo{journal}{Mol. Phys.} \textbf{\bibinfo{volume}{95}},
  \bibinfo{pages}{131} (\bibinfo{year}{1998}).

\bibitem[{\citenamefont{Vega}(1998)}]{V98}
\bibinfo{author}{\bibfnamefont{C.}~\bibnamefont{Vega}}, \bibinfo{journal}{J.
  Chem. Phys.} \textbf{\bibinfo{volume}{108}}, \bibinfo{pages}{3074}
  (\bibinfo{year}{1998}).

\bibitem[{\citenamefont{Tukur et~al.}(1999)\citenamefont{Tukur, Hamad, and
  Mansoori}}]{THM99}
\bibinfo{author}{\bibfnamefont{N.~M.} \bibnamefont{Tukur}},
  \bibinfo{author}{\bibfnamefont{E.~Z.} \bibnamefont{Hamad}}, \bibnamefont{and}
  \bibinfo{author}{\bibfnamefont{G.~A.} \bibnamefont{Mansoori}},
  \bibinfo{journal}{J. Chem. Phys.} \textbf{\bibinfo{volume}{110}},
  \bibinfo{pages}{3463} (\bibinfo{year}{1999}).

\bibitem[{\citenamefont{Regnaut et~al.}(2001)\citenamefont{Regnaut, Dyan, and
  Amokrane}}]{RDA01}
\bibinfo{author}{\bibfnamefont{C.}~\bibnamefont{Regnaut}},
  \bibinfo{author}{\bibfnamefont{A.}~\bibnamefont{Dyan}}, \bibnamefont{and}
  \bibinfo{author}{\bibfnamefont{S.}~\bibnamefont{Amokrane}},
  \bibinfo{journal}{Mol. Phys.} \textbf{\bibinfo{volume}{99}},
  \bibinfo{pages}{2055} (\bibinfo{year}{2001}).

\bibitem[{\citenamefont{Henderson}(1975)}]{H75}
\bibinfo{author}{\bibfnamefont{D.}~\bibnamefont{Henderson}},
  \bibinfo{journal}{Mol. Phys.} \textbf{\bibinfo{volume}{30}},
  \bibinfo{pages}{971} (\bibinfo{year}{1975}).

\bibitem[{\citenamefont{Santos et~al.}(2001)\citenamefont{Santos, Yuste, and
  {L\'opez de Haro}}}]{SYH01}
\bibinfo{author}{\bibfnamefont{A.}~\bibnamefont{Santos}},
  \bibinfo{author}{\bibfnamefont{S.~B.} \bibnamefont{Yuste}}, \bibnamefont{and}
  \bibinfo{author}{\bibfnamefont{M.}~\bibnamefont{{L\'opez de Haro}}},
  \bibinfo{journal}{Mol. Phys.} \textbf{\bibinfo{volume}{99}},
  \bibinfo{pages}{1959} (\bibinfo{year}{2001}).

\bibitem[{\citenamefont{Carnahan and Starling}(1969)}]{CS69}
\bibinfo{author}{\bibfnamefont{N.~F.} \bibnamefont{Carnahan}} \bibnamefont{and}
  \bibinfo{author}{\bibfnamefont{K.~E.} \bibnamefont{Starling}},
  \bibinfo{journal}{J. Chem. Phys.} \textbf{\bibinfo{volume}{51}},
  \bibinfo{pages}{635} (\bibinfo{year}{1969}).

\bibitem[{\citenamefont{Erpenbeck and Luban}(1985)}]{EL85}
\bibinfo{author}{\bibfnamefont{J.~J.} \bibnamefont{Erpenbeck}}
  \bibnamefont{and} \bibinfo{author}{\bibfnamefont{M.~J.} \bibnamefont{Luban}},
  \bibinfo{journal}{Phys. Rev. A} \textbf{\bibinfo{volume}{32}},
  \bibinfo{pages}{2920} (\bibinfo{year}{1985}).

\bibitem{BS01}
C. Barrio and J. R. Solana, J. Chem. Phys. \textbf{11}, 7123 (2001).
It should be pointed out that the expressions for $g_{ij}(\sigma_{ij})$ of hard-disk mixtures introduced in this paper were also
independently derived by Jenkins and Mancini. \cite{JM87}



\bibitem[{\citenamefont{Luding and Strau\ss}(2001)}]{LS01}
\bibinfo{author}{\bibfnamefont{S.}~\bibnamefont{Luding}} \bibnamefont{and}
  \bibinfo{author}{\bibfnamefont{O.}~\bibnamefont{Strau\ss}}, in
  \emph{\bibinfo{booktitle}{{Granular Gases}}}, edited by
  \bibinfo{editor}{\bibfnamefont{T.}~\bibnamefont{P{\"o}schel}}
  \bibnamefont{and} \bibinfo{editor}{\bibfnamefont{S.}~\bibnamefont{Luding}}
  (\bibinfo{publisher}{Springer}, \bibinfo{address}{Berlin},
  \bibinfo{year}{2001}), pp. \bibinfo{pages}{389--409}.

\bibitem[{\citenamefont{Malijevsk\'y et~al.}(1997)\citenamefont{Malijevsk\'y,
  Baro\v{s}ov\'a, and Smith}}]{MBS97}
\bibinfo{author}{\bibfnamefont{A.}~\bibnamefont{Malijevsk\'y}},
  \bibinfo{author}{\bibfnamefont{M.}~\bibnamefont{Baro\v{s}ov\'a}},
  \bibnamefont{and} \bibinfo{author}{\bibfnamefont{W.~R.} \bibnamefont{Smith}},
  \bibinfo{journal}{Mol. Phys.} \textbf{\bibinfo{volume}{91}},
  \bibinfo{pages}{65} (\bibinfo{year}{1997}).

\bibitem[{\citenamefont{Malijevsk\'y}(2002)}]{M02}
\bibinfo{author}{\bibfnamefont{A.}~\bibnamefont{Malijevsk\'y}},
  \bibinfo{howpublished}{private communication} (\bibinfo{year}{2002}).

\bibitem[{\citenamefont{Sear and Mulder}(1998)}]{SM98}
\bibinfo{author}{\bibfnamefont{R.}~\bibnamefont{Sear}} \bibnamefont{and}
  \bibinfo{author}{\bibfnamefont{B.~M.} \bibnamefont{Mulder}},
  \bibinfo{journal}{Mol. Phys.} \textbf{\bibinfo{volume}{93}},
  \bibinfo{pages}{181} (\bibinfo{year}{1998}).

\bibitem[{\citenamefont{Frisch and Percus}(1999)}]{FP99}
\bibinfo{author}{\bibfnamefont{H.~L.} \bibnamefont{Frisch}} \bibnamefont{and}
  \bibinfo{author}{\bibfnamefont{J.~K.} \bibnamefont{Percus}},
  \bibinfo{journal}{Phys. Rev. E} \textbf{\bibinfo{volume}{60}},
  \bibinfo{pages}{2942} (\bibinfo{year}{1999}).

\bibitem[{\citenamefont{Bishop et~al.}(1999)\citenamefont{Bishop, Masters, and
  Clarke}}]{BMC99}
\bibinfo{author}{\bibfnamefont{M.}~\bibnamefont{Bishop}},
  \bibinfo{author}{\bibfnamefont{A.}~\bibnamefont{Masters}}, \bibnamefont{and}
  \bibinfo{author}{\bibfnamefont{J.~H.~R.} \bibnamefont{Clarke}},
  \bibinfo{journal}{J. Chem. Phys.} \textbf{\bibinfo{volume}{110}},
  \bibinfo{pages}{11449} (\bibinfo{year}{1999}).

\bibitem[{\citenamefont{Parisi and Slanina}(2000)}]{PS00}
\bibinfo{author}{\bibfnamefont{G.}~\bibnamefont{Parisi}} \bibnamefont{and}
  \bibinfo{author}{\bibfnamefont{F.}~\bibnamefont{Slanina}},
  \bibinfo{journal}{Phys. Rev. E} \textbf{\bibinfo{volume}{62}},
  \bibinfo{pages}{6554} (\bibinfo{year}{2000}).

\bibitem[{\citenamefont{Yuste et~al.}(2000)\citenamefont{Yuste, Santos, and
  {L\'opez de Haro}}}]{YSH00}
\bibinfo{author}{\bibfnamefont{S.~B.} \bibnamefont{Yuste}},
  \bibinfo{author}{\bibfnamefont{A.}~\bibnamefont{Santos}}, \bibnamefont{and}
  \bibinfo{author}{\bibfnamefont{M.}~\bibnamefont{{L\'opez de Haro}}},
  \bibinfo{journal}{Europhys. Lett.} \textbf{\bibinfo{volume}{52}},
  \bibinfo{pages}{158} (\bibinfo{year}{2000}).

\bibitem{FSL02}
R. Finken, M. Schmidt, and H. L\"owen, Phys. Rev. E \textbf{65}, 016108 (2002).

\bibitem[{\citenamefont{Enciso et~al.}(2002)\citenamefont{Enciso, Almarza,
  Gonz\'alez, and Bermejo}}]{EAGB02}
\bibinfo{author}{\bibfnamefont{E.}~\bibnamefont{Enciso}},
  \bibinfo{author}{\bibfnamefont{N.~G.} \bibnamefont{Almarza}},
  \bibinfo{author}{\bibfnamefont{M.~A.} \bibnamefont{Gonz\'alez}},
  \bibnamefont{and} \bibinfo{author}{\bibfnamefont{F.~J.}
  \bibnamefont{Bermejo}}, \emph{\bibinfo{title}{{The virial coefficients of
  hard hypersphere binary mixtures}}}, \bibinfo{howpublished}{Mol. Phys., in
  press} (\bibinfo{year}{2002}).

\bibitem[{\citenamefont{Luban and Michels}(1990)}]{LM90}
\bibinfo{author}{\bibfnamefont{M.}~\bibnamefont{Luban}} \bibnamefont{and}
  \bibinfo{author}{\bibfnamefont{J.~P.~J.} \bibnamefont{Michels}},
  \bibinfo{journal}{Phys. Rev. A} \textbf{\bibinfo{volume}{41}},
  \bibinfo{pages}{6796} (\bibinfo{year}{1990}).

\bibitem[{\citenamefont{Baro\v{s}ov\'a
  et~al.}(1996)\citenamefont{Baro\v{s}ov\'a, Malijevsk\'y, Lab\'{\i}k, and
  Smith}}]{BMLS96}
\bibinfo{author}{\bibfnamefont{M.}~\bibnamefont{Baro\v{s}ov\'a}},
  \bibinfo{author}{\bibfnamefont{A.}~\bibnamefont{Malijevsk\'y}},
  \bibinfo{author}{\bibfnamefont{S.}~\bibnamefont{Lab\'{\i}k}},
  \bibnamefont{and} \bibinfo{author}{\bibfnamefont{W.~R.} \bibnamefont{Smith}},
  \bibinfo{journal}{Mol. Phys.} \textbf{\bibinfo{volume}{87}},
  \bibinfo{pages}{423} (\bibinfo{year}{1996}).

\bibitem[{\citenamefont{Sanchez}(1994)}]{S94}
\bibinfo{author}{\bibfnamefont{I.~C.} \bibnamefont{Sanchez}},
  \bibinfo{journal}{J. Chem. Phys.} \textbf{\bibinfo{volume}{101}},
  \bibinfo{pages}{7003} (\bibinfo{year}{1994}).

\end{thebibliography}

\end{document}